\documentclass[12pt,a4paper]{article}
\usepackage[T1]{fontenc}
\usepackage{epsf}
\usepackage{setspace}
\date{}
\begin{document}

\title{Jets with Reversing Buoyancy}
\author{R.~V.~R.~Pandya\thanks{rvrptur@yahoo.com} \\[1em] and 
P.~Stansell\thanks{paulstansell@gmail.com}}
\maketitle

\renewcommand{\thefootnote}{\fnsymbol{footnote}}

\begin{abstract}
A jet of heavy fluid is injected upwards, at time $t=0$, into a lighter
fluid and reaches a maximum height at time $t=t_i$ and then flows back
around the upward flow.  A similar flow situation occurs for a light
fluid injected downward into a heavy one.  In this paper an exact
analytical expression for $t_i$ is derived.  The expression remains
valid for laminar and turbulent buoyant jets with or without swirl. 
\end{abstract}
\thispagestyle{empty}

\section{Introduction}

The study of jets with reversing buoyancy has found many applications in
engineering.  A heavy fluid jet injected upward into a lighter fluid is
referred to as a negatively buoyant jet or fountain.  This type of flow
situation often occurs in disposal of industrial effluent.  Typical
examples are brine and gypsum discharged into the ocean through
multi-port diffusers. 

Similarly, a positively buoyant jet occurs when a light fluid is
injected downwards into a heavier one.  This jet reaches a maximum depth
and then turns upward and rises around the downward flow.  These types
of jets are also referred to as inverted fountains.  Typical examples
are heating of a large open structure by fan-driven heaters on the
ceiling and mixing of a two-layer water reservoir with propellers
\cite{Baines_1}.  In both cases the jets also possess some swirl. 

One of the first experimental studies on negatively buoyant jets was
conducted by Turner \cite{Turner}.  He used a nozzle to inject a salt
solution of density $\rho_0$ upward into a tank of stationary fresh
water of density $\rho_a < \rho_0$.  He measured the maximum height,
$z_i$, attained by the jet and the mean height, $z_m < z_i$, at which
the jet finally settles down.  He related the two parameters $z_i/D$ and
$z_m/D$ as proportional to the densimetric Froude number
\[
F_r = \left( \frac{U_0^2 \, \rho_0}{g \, D \, (\rho_0-\rho_a)}
\right)^\frac{1}{2}
\]
by using a dimensional analysis.  Here, $U_0$ is the uniform velocity at
the nozzle outlet, $g$ is the acceleration due to gravity and $D$ is
diameter of the nozzle.

Later, Abraham \cite{Abraham} studied the problem theoretically and
confirmed the result of Turner.  Recently Baines \emph{et al.\@}
\cite{Baines_1,Baines_2} made an extensive study, both experimentally
and theoretically on negatively buoyant jets.  None of these studies
provide an expression for $t_i$.  Here a simple analysis is presented to
obtain an exact analytical expression for $t_i$.

\section{Analysis}

\begin{figure}
\unitlength=1mm
\centering
\begin{minipage}{80mm}
\setlength{\epsfxsize}{80mm}
\epsffile{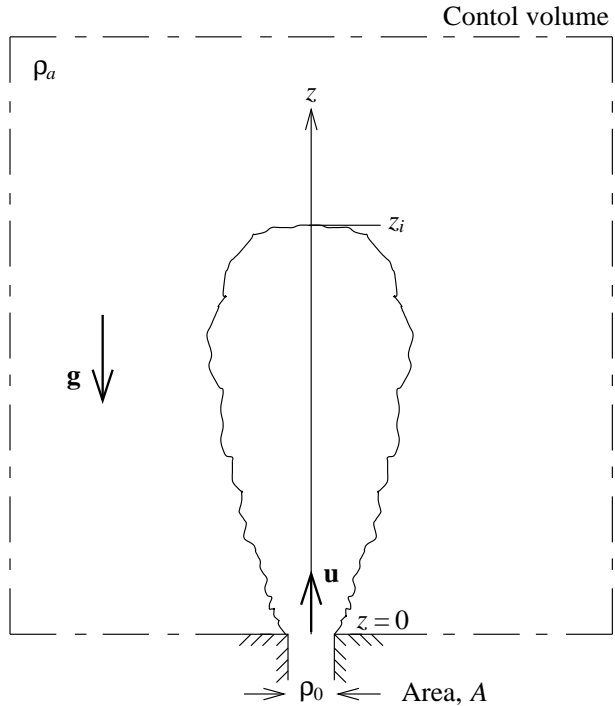}
\end{minipage}
\caption{A schematic diagram of a negative buoyancy jet at $t=t_i$.}
\label{fig:jet}
\end{figure}

A schematic diagram of the jet and control volume at time $t=t_i$ is 
shown in Figure~\ref{fig:jet}.

Consider a flow situation in which a fluid of density $\rho_0$ is
injected, at time $t=0$ in the upward direction at a constant flow rate,
from a nozzle at $z=0$ into an ambient fluid of density $\rho_a <
\rho_0$.  Let the cross sectional area of the nozzle be $A$ and the
fluid velocity distribution at the nozzle to be $u$.  The jet
continuously looses its momentum due to the reverse buoyancy force and
reaches a maximum height $z_i$ at time $t=t_i$ after which the jet turns
downwards.

For times $t<t_i$ the $z$-component of the momentum in the control
volume increases continuously until time $t=t_i$ from which time it
decreases due to the downward flow.  Therefore, at $t=t_i$, the rate at
which the momentum is changing inside the control volume is equal to
zero.  Application of the momentum theorem to the flow situation at time
$t=t_i$ implies that the net force ${\bf F}_b$ (in this case the
buoyancy force in the $z$-direction) acting on the control volume is
equal to the rate at which the net momentum ${\bf M}$ is leaving the
control volume, that is to say,
\begin{equation}
{\bf F}_b = {\bf M}
\label{eq:definition of F}
\end{equation}
If we take the control volume to be sufficiently large then the net rate
at which the momentum is leaving the control volume can be written as
\begin{equation}
{\bf M} = - \rho_0 \, \int_A u^2 \, d A' \, {\bf k},
\label{eq:value of M}
\end{equation}
where ${\bf k}$ is the unit vector in the $z$-direction.

The total volume of the fluid jet of density $\rho_0$ entering the
control volume in time $t_i$ is
\[
V = t_i \, \int_A u \, d A'.
\]
This is also equal to the volume of fluid of density $\rho_a$ displaced
by the jet in time $t_i$.  Therefore, the net buoyancy force acting on
the fluid inside the control volume is
\begin{equation}
{\bf F}_b = - t_i \, (\rho_0 -\rho_a) \, g \, \int_A u \, d A' \, {\bf
k}. 
\label{eq:value of F}
\end{equation}

From equations~(\ref{eq:definition of F}), (\ref{eq:value of M}) and
(\ref{eq:value of F}) we obtain an analytic expression for $t_i$ as
\begin{equation}
t_i = \frac{\rho_0 \int_A u^2 \, d A'}{(\rho_0 -\rho_a) \, g\, \int_A u
\, d A'},
\label{eq:value for t_i}
\end{equation}
which reduces to
\begin{equation}
t_i = \frac{U_0 \, \rho_0}{|\rho_0 -\rho_a| \, g},
\label{eq:t_i modulus}
\end{equation}
for $u=U_0=\mbox{constant}$ over the area $A$.  The modulus is
introduced so that equation~(\ref{eq:t_i modulus}) is valid for
positively and negatively buoyant jets.  It is worth mentioning that we
have not made any assumptions about the regime of flow.  Therefore the
expression for $t_i$ remains valid for laminar and turbulent buoyant
jets with reversing buoyancy.  Also, as any swirl added to the fluid at
the nozzle does not affect the $z$-momentum of the fluid,
equation~(\ref{eq:value for t_i}) remains valid for this case too.

\section{Notation}

The following symbols are used in this paper:
\vspace{-1em}
\begin{tabbing}
symbol \= definition \kill \\
$t$ \> time \\
$t_i$ \> time at which the jet reaches its maximum height \\
$\rho_0$ \> density of the fluid in jet \\
$\rho_a$ \> density of the ambient fluid \\
$z_i$ \> maximum height attained by the jet \\
$z_m$ \> mean height at which the jets settles \\
$F_r$ \> Froude number \\
$U_0$ \> uniform fluid velocity as the nozzle outlet \\
$D$ \> diameter of the nozzle outlet\\
$g$ \> acceleration due to gravity \\
$A$ \> cross sectional area of nozzle outlet \\
$u$ \> fluid velocity distribution at the nozzle outlet \\
${\bf F}_b$ \> buoyancy force \\
${\bf M}_b$ \> rate at which momentum is leaving the control volume \\
${\bf M}_b$ \> rate at which momentum is leaving the control volume \\
${\bf k}$ \> unit vector in the $z$-direction \\
$V$ \> total volume of jet fluid
\end{tabbing}

\end{document}